We suggest that by virtue of the very early processes immediately following the big bang incipient $1^{st}$ generation particles are created forming periodic structures (strings) to confine the quarks. These strings may be described either by the Laplace equation, due to globally vanishing space charges, or by Poisson-Boltzmann equation, due to finite charges. We also propose that the supersymmetry created by the big bang is broken through extended vibronic mixing at the end of the $0^{th}$ generation giving rise to the diversity of atoms and molecules at later stages of the evolving universe. We choose quark confinement by electrostatic strings, mostly those supplied by 2D solutions of the PB equatiom. The possible role of $ML_n$ molecule appearing as a further molecular generation is also discussed.

1. Introduction

The properties of all the elementary particles (twelve fermions and four bosons) known toward this date are fairly well described by the Standard Model (SM) [1]. Yet, there are two notable exceptions: the gravitons (gravitational quanta) and the Higgs boson. Of these, observing the latter is of key importance for the model with a high rate of expectancy at the LHC machine, while the former is an enigma not expected to be resolved any time soon.

The fermions (matter particles) are classified into three generations as regards their expected appearance relative to the big bang (bb). The bosons (force particles) mediate the interaction between fermions. The Higgs mechanism is believed to give masses at rest to all the fundamental particles except for the electromagnetic force massless photon and the strong force massless gluon but including the finite mass W and Z weak force bosons.

Several SM virtues are commonly recognized: SM predicted the existence of the W and Z bosons, the gluon, and the top and charm quarks before these particles were observed. Their predicted properties were later experimentally confirmed with a good precision. The success of SM is evidenced by the measured masses of the W and Z bosons comparing favorably with SM predicted masses.

Yet, there are some challenges too, such as the Higgs boson not yet observed, the reported finite neutrino mass, etc. The model tells little if anything of the events preceding the $1^{st}$ generation and, for that matter, scarcely about the timing of supersymmetry and its possible breakup at some stage of the generation.

The material downside is distributed into 6 sections. Sections 1 through 3 deal with solutions of dimensions 1D through 3D, respectively. Section 4 is an introduction to the $ML_n$ molecule, stressing on properties that may be found useful for our purpose, which is the geometry of bb projectiles. Do not forget that in accordance with the general relativity, particles themselves build up the space-time curvature. Some brief debate is further offered to the bb plausible scenario in part 5 and very little, if any, to the color confinement by electrostatic strings in part 6 under *discussion*.

2. 1D (planar) Coulomb lattice gas models

2.1. Boltzmann-tail statistics for classical particles

Hereby we do not attach any global character to supersymmetry. We rather consider it having an extended local feature. As an example for the extended feature, we take a quadratic planar lattice composed of alternating positive and negative charges. We choose the simple planar quadrate, since the solutions of both Laplace (L) [2] and Poisson-Boltzmann (PB) [3] equations are known for this structure which can be compared with each other and the respective free energies derived to seek the thermal equilibrium [4]. Three pair equilibria have been sought: $F_L \leftrightarrow F_{random\ phase}$, $F_{PB} \leftrightarrow F_{random\ phase}$, and $F_L \leftrightarrow F_{PB}$. Under the electro-neutrality assumption the negative and positive charges compensate each other. Each boson particle has a fermion superpartner and vice versa, as required by the supersymmetry.

Examples of free energy calculations are as shown in Figure 1. A system of positive and negative charges created instantly by the bb at t = 0 are distributed to form a wavelike structure of either L or PB type as described by the following equations [4]:

$$n_\pm(\mathbf{r}) \sim n_0 \exp(\pm e\varphi(\mathbf{r})/k_BT) \tag{1}$$

holding good for Boltzmann-tail statistics with $n_0 = \exp[-(\mu - \varepsilon)/(k_BT)]\}$ ($\mu$ is the chemical potential or Fermi energy) to determine the bulk charge, if any, while the free energy is:

$$F = \int d\mathbf{r}\ [F^+n^+ + F^-n^- + (F^+ + F^- - B_P)n_P] - TS$$

$$S = -k_B \int d\mathbf{r}\{n^+[\ln(n^+/N) - 1] + n^-[\ln(n^-/N) - 1] + n^P[\ln(n^P/z^P) - 1]\}$$

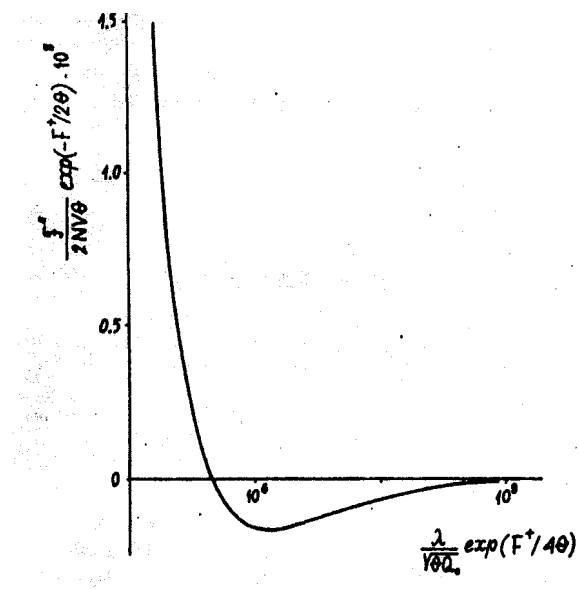

Figure 1: Free energy of a 1D planar PB structure in appropriate coordinates, after Ref. [4].

Here S is the configurational entropy at small particle densities (n « N), $F^{+-}$ are the formation free energies of the ± particles of concentrations $n_\pm$, respectively, $B_P$ and $n_P$ are binding energy and concentration of pairs, $z_P$ is the coordination number, if any pairs form at the later stages of the sea of fundamental particles. Minimizing eqn. (1) with respect to the particle densities we get

$$n_\pm = N\exp[-(F^\pm - e\varphi)/k_BT]$$
$$n_P = Nz_P\exp[-(F^+ + F^- - B_P)/k_BT] \tag{2}$$

at thermal equilibrium under constant temperature T and pressure p. Here $\varphi = \varphi(\mathbf{r})$ is the electrostatic potential obtainable by solving the PB equation at bulk charge density

$$\rho(\mathbf{r}) = -2e\ \tilde{N}\ \sinh(\psi) \tag{3}$$

$$\tilde{N} = N\exp[-(F^+ - e\varphi_\infty)/k_BT]$$

$$e\varphi_\infty/k_BT = \exp[(F^+ - F^-)/2k_BT]$$

$$\psi = (e\varphi - e\varphi_\infty)/k_BT. \tag{4}$$

$$\Delta\psi = \upsilon^2 \sinh\psi \tag{5}$$

is the resulting Poisson-Boltzmann equation. Equivalently, we get the PB equation in reduced form appropriate for the 1D planar symmetry

$$d^2\psi/dZ^2 = \sinh\psi \tag{5'}$$

with X,Y,Z = $\upsilon$ x,y,z with $\upsilon = r_D^{-1} = \sqrt{(8\pi Ne^2/\kappa k_BT)}$, the reciprocal Debye screening radius, $\kappa$ is the dielectric constant, $\Delta = d^2/dZ^2$. Here and above N stands for the total number of sites in the sea of fundamental particles. The quantity $\psi$ is the reduced potential relative to the 'vanishing charge potential' $\varphi_\infty$. In 1D, the solutions are either periodic or aperiodic functions:

$$\psi(\upsilon z) = \ln\cotan(\tfrac{1}{2}\upsilon z)\quad \text{periodic}$$
$$\psi(\upsilon z) = \ln\tanh(\tfrac{1}{2}\upsilon z),\ \text{etc.}\quad \text{aperiodic,} \tag{5''}$$

as well as a whole variety of solutions in between. We stress that by definition the PB equation is meaningful at thermal equilibrium only in so far as the bulk density ρ is given by the statistical charge distribution within the medium and can be defined at whatever classical particle statistics from Boltzmann-tail statistics through Fermi statistics with comparable justification for the resulting PB equation [5,6]. This will be shown below. The preferences for the one or the other statistics will be discussed at some length shortly. For fermions generated by the bb, Fermi-Dirac's statistics will be compulsory, as will Bose-Einstein's statistics for bosons. Nevertheless, we start by using Boltzmann-tail statistics due to its educational and even heuristic merits at the later stages of the bb excitation. Moreover, it is well known that fermions behave classically at energies well above the Fermi energy.

2.2. Fermi-Dirac statistics for matter particles

The statistical equilibrium issue has already been dealt with in a plane-wave context [4,6]. Both Fermi-Dirac and Bose-Einstein statistics have been considered lately [7]. On using the complete form of the configurational enthropy terms (the small n condition no longer valid), the resulting particle densities are:

$n_{\pm} = \underline{N} / [P + \exp(\pm \psi)]$ (6)

with $P = \exp[-(F^+ - e\varphi_\infty)/k_BT]$ and $\underline{N} = NP$. Inserting into the Poisson equation leads to

$\Delta\psi = \upsilon^2 \sinh\psi / (1 + P^2 + 2P \cosh\psi)$ (7)

Equation (6) reminds of the Fermi distribution. We call it Poisson-Fermi equation. As before in Section 2, the entire accessible volume has been divided into N sub-volumes, to each one an occupation number 0 or 1 being attached. The statistics in (6) passes into Boltzmann's (tail) statistics for P « 1, that is, for T » $(e\varphi_\infty - F^+)/k_B$. It is remarkable that Fermi's statistics appears in a natural way as long as the complete forms of the configurational entropy equation (2) are invoked. This implies that the Fermi-Dirac law is inherent for our system, as are the derivative equations of the Boltzmann-tail statistics. The statistics benchmark is the average particle number [8,9]:

$<n_k> = \sum_{nk} n_k P(E_\alpha)_{N\alpha} = u_k / (1 + u_k) = 1 / \{ \exp[(\varepsilon_k - \mu) / k_BT] + 1 \}$ (8)

It might be argued that incipient fermions immediately following the bb obey Fermi-Dirac's statistics though due to the immense temperatures this could rather be the statistics' classical Boltzmann- tail form. Consequently, classical behavior of matter particles might be wide spread at the early stages after the big bang. This may not be the case for the force-mediating particles. The immediate consequence is that the exclusion principle may not be so restrictive initially as it will eventually turn out later. Ultimately, this will eventually create very favorable conditions under thermal equilibrium for matter build up after a time-delay.

2.3. Bose-Einstein statistics for force mediator particles

The 1D pattern described above presents the particle and related electrostatic distributions along an axis perpendicular to the equipotential planes of size largely superior to the Debye length $r_D$. So far we have done this for classic particles (fermions at energies above the sea of fundamental particle energies), as well as for fermions (matter particles) under the Fermi energy. The resulting PB equations have been shown solvable in both periodic and aperiodic functions. It is tempting to see how the force mediators behave under equilibrium conditions following the bb. This can be done by manipulating the sign of the vanishing charge potential, as it follows from $\underline{N} = N\exp[-(F^+ - e\varphi_\infty)/k_BT] = N\exp[-(F^+ + F^-)/2k_BT]$.

The statistics benchmark is again the definition of an average number of particles [9]:

$$<n_k> = \sum_{nk} n_k P(E_n)_{N\alpha} = u_k / (1 - u_k) = 1 / \{ \exp[(\varepsilon_k - \mu) / k_BT] - 1 \} \qquad (9)$$

To abide by Bose-Einstein's statistics one should change the sign to P in equation (7) and compare the result with (9). However, changing the sign to P would imply introducing imaginary potentials. Consequently, while Fermi–Dirac's statistics seems inherent for lattice gas particles, Bose–Einstein's statistics does not. This is an interesting case where matter particles A and B coexist faintly with force mediators C and pair interactions between the former two do not occur outright at the $0^{th}$ generation stage. It may be argued that boson superpartners may exist in kind of a fluid state within the 1D periodic structure with matter fermions occupying stable positions, as in a real metal where conduction particles (electrons) form a gas or fluid state within the space provided by the residual ions. (The actual identity of particles, conduction and residual, as electrons and ions, respectively, is opposite to the one of our hypothetical 1D structure of the $0^{th}$ generation, though metals occur at much later stages of the universe.) Alternatively, bosons may form by fermions as Cooper or real-space pairs.

Actually, force mediators should not be anticipated to building a lattice, though they would certainly be expected to promote one. An example will be provided at the later stages of a fully blown universe, as electromagnetic interactions promoted by photons (bosons) stimulate the formation of an ionic crystalline lattice composed of positive and negative ions through electrostatic interactions.

3. 2D&3D Coulomb lattice gas models

Most of the early work on the equations of electrostatics have concentrated on Laplace's equation, due to its application in crystallography. Some of the scientific interest has moved to the Poisson-Boltzmann equations because of increased appreciation of their significance [9]. Nevertheless, some doubts remain unsolved as to the PB statistical consistency [10].

3.1. Laplace equation

3D periodic solutions to the Laplace equation have been reported more recently [10]. Historically, solutions to Laplace's equation have usually been found by separating the variables; however, the method is short of providing 3D solutions derived differently

$$\psi(\mathbf{r}) = \ln[(1 + \lambda)/(1 - \lambda)]^2 \qquad (10)$$

$\lambda = Au(\alpha x, k_1)v(\beta y, k_2)t(\gamma z, k_3) + B\underline{u}(\alpha x, k_1)\underline{v}(\beta y, k_2)\underline{t}(\gamma z, k_3) + C\, u^*(\alpha x, k_1)v^*(\beta y, k_2)\, t^*(\gamma z, k_3)$

$u(\alpha x, k_1)$, etc. are some of 12 generated elliptic functions, $k_i$ are the elliptic integral moduli.

3.2. Poisson-Boltzmann equations

Following an extensive study of the PB equations in 1D, including solutions and equilibrium with a random phase distributions, work has been done aimed at extending these premises to higher dimensions such as 2D and 3D. First, Martinov and Ouroushev [11] have presented arguments of the tendency for the formation of 2D self-consistent structures in Coulomb systems thus opening a line of good papers on the matter. The authors point out that the nonlinear trends may only be revealed when using the complete PB equation rather than its linearized Debye form. In so far as the latter form is obtained assuming $\psi \ll 1$ only and no other binding restrictions on the parameters in (5), one may expect the nonlinear solutions to exhibit non-linearities, as they actually occur at the charge accumulation sites (ionic islands). Without going into the details, the authors show the generating equations of the Jacobi elliptic functions emerging from the consequent manipulations, thereby indicating that the nonlinear periodicity is expressible in elliptic functions. They also deal with the character of the potential functions around the ionic sites to show that the singularities are typical point charge in character. An example is given by the nonlinear periodic function [12]:

$$\psi = 4\tanh^{-1}[cn(\alpha x, k_1) cn(\alpha y, k_2)], \qquad k_1^2 + k_2^2 = 1 \qquad (11)$$

which is found to solve the 2D PB equation.

Unlike the 1D planar configurations, the 3D solutions depict crystal-like structures exhibiting ionic sites and space charge regions in between. The latter peculiarity distinguishes the 3D PB feature from 3D Laplace structures, where the interionic space is empty. Nevertheless Laplace structures have found applications in various problems of electrostatics.

4. Symmetrized displacements of normal coordinates for $ML_n$ molecule

In addition to PB's electrostatic strings, $ML_n$ molecules merit an increased attention as bb products arising at later stages of the universe beyond SM. Below, we summarize symmetrized coordinates and transformation properties of some normal modes of the tetrahedral and octahedral representations for certain molecules. A complete list may be found in reference [13] and references cited therein. These examples of molecular systems are actual and have been widely used and discussed in relation with the dynamic processes in high-temperature superconductors and colossal magnetoresistance materials. For this reason we consider them more appropriate for our immediate purpose, as we place the bb happening at $ML_n$ and follow the temporal development of its products along the radial routes.

It is tempting to attribute the grand matter creating explosion to one molecule of whatever symmetry placed at the center of the incipient space at M while the protuberances of matter going out along the radial directions. In as much as solutions for both $LM_4$ and $LM_6$ are known, we choose either of them, perhaps with some preference for the latter, the grand molecule. Its excited form $ML_6^*$ may have given birth to the universe.

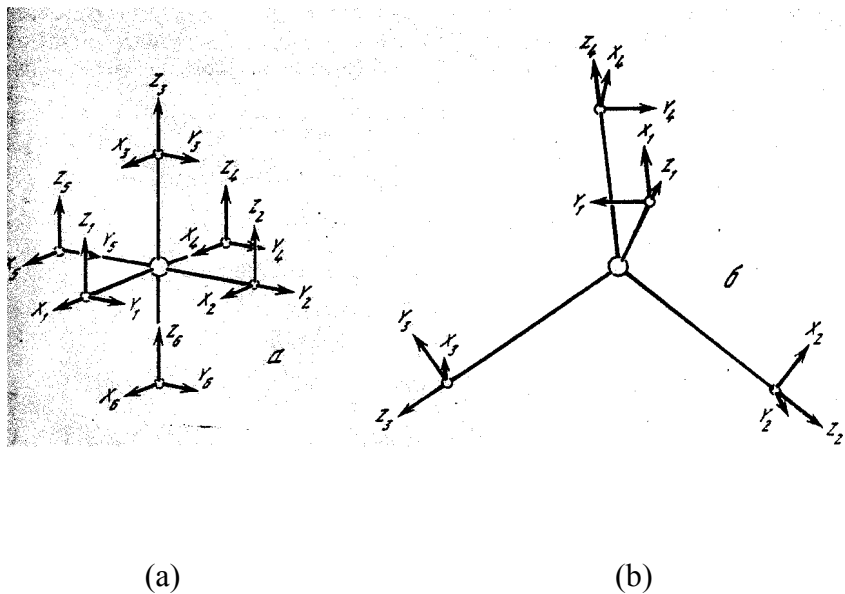

(a) (b)

Figure 2: General and local coordinate systems for octahedral $ML_6$ (a) and tetrahedral $ML_4$ (b) molecules [13]. One of these may be expected to be the progenitor of the universe.

The following tables summarize the symmetrized displacements and transformations of the normal coordinates of the tetrahedral and octahedral molecules [13].

Table I

Tetrahedral molecular system $ML_4$ of symmetry $T_d$

| Displacement | Symmetry type | Transformation | Fig. 5 coordinates exressed in Cartesian coordinates |
|---|---|---|---|
| $Q_\alpha$ | A | $x^2 + y^2 + z^2$ | ½ $(Z_1 + Z_2 + Z_3 + Z_4)$ |
| $Q_\theta$ | E | $2z^2 - x^2 - y^2$ | ½ $(X_1 - X_2 - X_3 + X_4)$ |
| $Q_\varepsilon$ | " | $x^2 - y^2$ | ½ $(Y_1 - Y_2 - Y_3 + Y_4)$ |
| $Q_\xi$ | $T_2'$ | $yz$ | ½ $(Z_1 - Z_2 + Z_3 - Z_4)$ |
| $Q_\eta$ | " | $zx$ | ½ $(Z_1 + Z_2 - Z_3 - Z_4)$ |
| $Q_\zeta$ | " | $xy$ | ½ $(Z_1 - Z_2 - Z_3 + Z_4)$ |

Table II

Octahedral molecular system $ML_6$ of symmetry $O_h$

| Displacement | Symmetry type | Transformation | Fig. 5 coordinates exressed in Cartesian coordinates |
|---|---|---|---|
| $Q_\alpha$ | $A_{1g}$ | $x^2 + y^2 + z^2$ | $(X_2 - X_5 + Y_3 - Y_6 + Z_1 - Z_4)/\sqrt{6}$ |
| $Q_\theta$ | $E_g$ | $2z^2 - x^2 - y^2$ | $(2Z_3 - 2Z_6 - X_1 + X_4 - Y_2 + Y_5)/\sqrt{3}$ |
| $Q_\varepsilon$ | " | $x^2 - y^2$ | $½ (X_1 - X_4 - Y_2 + Y_5)$ |
| $Q_\xi$ | $T_{2g}$ | $yz$ | $½ (Z_2 - Z_5 + Y_3 - Y_6)$ |
| $Q_\eta$ | " | $zx$ | $½ (X_3 - X_6 + Z_1 - Z_4)$ |
| $Q_\zeta$ | " | $xy$ | $½ (Y_1 - Y_4 + X_2 - X_5)$ |
| $Q_x'$ | $T_{1u}'$ | $x$ | $½ (X_2 + X_3 + X_5 + X_6)$ |
| $Q_y'$ | " | $y$ | $½ (Y_1 + Y_2 + Y_4 + Y_6)$ |
| $Q_z'$ | " | $z$ | $½ (Z_1 + Z_2 + Z_4 + Z_5)$ |
| $Q_x''$ | $T_{1u}''$ | $x$ | $½ (X_1 + X_4) / \sqrt{2}$ |
| $Q_y''$ | " | $y$ | $½ (Y_2 + Y_5) / \sqrt{2}$ |
| $Q_z''$ | " | $z$ | $½ (Z_3 + Z_6) / \sqrt{2}$ |

5. Plausible scenario

Being located at the "central point" of the pre-universe, whatever meaning you may ascribe to it, the grand progenitor $ML_6$ molecule is also the focal point of any disturbances. One or a few of them manage to excite the molecule to the extent of explosion at $t = 0$ (the big bang). Consequent product created in the form of fundamental matter and force particles flies out as protuberances along the symmetry axes of the molecule.

It seems likely that the physical law to control the outburst has something to do with the Poisson-Boltzmann equation which combines the basic law of electrostatics with the Boltzmann-tail statistics in view of the enormously harsh conditions (high temperatures). Consequently the simplified picture of the universe is an excited form of the progenitor molecule with beams of fundamental particles along its symmetry axes.

This is the picture during the $0^{th}$ stage which is missing in the standard model. One way or the other, the products of that stage are "incipient fundamental particles" bound to the symmetry axes of the grand molecule. This binding distinguishes them from the "genuine fundamental particles" of the subsequent $1^{st}$ stage of the generation. Once the universe has come to the $1^{st}$ stage we enter what the standard model has been talking about.

The preliminary $0^{th}$ generation stage should also be related to the supersymmetry breaking before the system has entered the broken symmetry stages $1^{st}$ to $3^{rd}$. This is the breakup of fermion – boson partnership and may be the prerequisite for the formation of stable molecules vital for life on earth. We have suggested that extended Jahn-Teller mixing of fermion states by bosons will create new fermion and boson states which may lift the supersymmetry [14].

6. Discussion

The foregoing scenario is controversial It rests on no specific calculations or experimental results. The sole justification may be found in the electrostatic interactions as described by PB and Laplace equations as they stand in 1D or higher symmetries. A tremendous effort has been spent to put the house in order: the derivation and classification of solutions in 1D, and, in particular, the periodic solutions in higher 2D and 3D symmetries which resemble crystal-like structures. This sort out work is still going on for the higher symmetries where it is by no means proven that all classes of solutions have been exhausted by the ones that have already been found. The relationship of periodic solutions, whether Laplace or Poisson-Boltzmann, with the crystalline lattices is obvious: the projection along the z-axis of a 1D structure is a one-dimensional lattice, the projection of a 2D structure in (x,y) plane is a two-dimensional lattice, and so far there are only 3D Laplace solutions known to represent a three-dimensional crystal lattice. At this point we may also add the, still not available, 3D periodic structures by PB solutions leading to three dimensional crystalline lattices. The relationship of all these periodic arrangements with real crystalline structures may be checked by comparing them with experimental crystallographic data.

To conclude our scenic proposals, we see that following the big bang matter particles can bind in 1D, 2D, or 3D Laplace or PB crystal-like arrangements or $ML_n$ molecules making them less free to migrate in the widening space around. This (quark) confinement is characteristic of the $0^{th}$ generation stage and is possibly removed at the beginning of the $1^{st}$ stage when fundamental matter and force particles become all free.

Further on, the opposition to the singular character of the "ionic islands" merits a special place though it does not pose any serious challenge. As a matter of fact, it has its origin in the point charge character of the ionic islands so that they are simply removable by cutoff [3]. For this reason we do not consider it justified to discard a contribution simply because it has features common for electrostatics. The so-called "statistical inconsistency" of the solutions of the complete PB equation are of related origin. Nevertheless, they have widely been used for solving physical problems here and there after applying a cutoff [15].

Finally, a related approach to post bb phenomena has been proposed by A.M. Polyakov some 35 years before based on the PB equation [16]. The author addresses the search for an universality class of (confining) strings for the quark confinement [17]. The strings are described by periodic solutions of the PB equation. In particular the lowest free energy periodic solution (5") has been applied to represent the 1D periodic arrangement of strings. No possible applications of other symmetries or free energies have been discussed. Nevertheless, the 2D picture following equation (11) may be more appropriate with strings extending along the z-axis. The advantage of using the 2D (x,y) solution over the 1D function is that the strings are z-independent along their length, whereas the motion is in (x,y) plane.

The picture just described is typical for the $0^{th}$ generation stage. Near the end of that stage all the quarks, and consequently, particles are confined with almost no chance of escaping, because of being driven back by the (electrostatic) strings. At the beginning of the $1^{st}$ stage, however, there appears a gradually growing flow of outgoing particles, due to a weakening string repel. During the subsequent three stages the trend is preserved and nearly all the quarks and particles become free to move unconfined.

Lately, we have suggested that extended Jahn-Teller processes may play a role in the breaking of supersymmetry soon after the big bang [14]. This may lead to the consumption of some force- mediating bosons over their matter superpartners to destroy the mutual balance. This would make regarding the extended vibronic coupling an important ingredient of any super-symmetry breaking theory.

The $ML_n$ molecules have lately become matter of interest for solid state physics after their role in high-$T_C$ superconductivity and colossal magnetoresistance has been appreciated. Both phenomena are kind of exotic appearances with the essential participation of vibronic polarons. In a way, the colossal magnetoresistances may be regarded as the fingerprints of vibronic polarons [18,19]. Below some transition temperature, say $T_T$, the vibronic polarons are free to move and carry an electric current which is largely decreased above $T_T$. The decrement is due to vibronic polaron confinement as the polarons arrange to form an insulating crystalline lattice which is nonconductive. In any event, this polaron confinement reminds of the quark confinement as discussed above and which has a similar effect on the matter particles since their color quarks are all confined or ultimately immobilized within the $0^{th}$ stage. Now, as the color quarks are confined, so are the corresponding matter particles (leptons) which cannot survive without the quarks.

As far as the high $T_C$ superconductivity is concerned, $CuO_6$ is the basic ingredient, the copper oxide octahedron, distorted by a JT elongation along the c-axis. The planar $Cu^{2+}.4(O^{2-})$ frame holding the conductive carriers as part of the octahedron has the $CuO_2$ composition. The vibronic polarons are somewhat less popular for high-$T_C$ materials though certain related units are not: these are bond polarons which conduct the interlayer currents between adjacent $CuO_2$ planes across the apical oxygens, as seen elsewhere [20]. The interlayer coupling is essential for the superconductivity.

Too little if anything is clear about the composition of the "grand progenitor molecule" which gives the starting material for the big bang and thereby for the birth of our present universe. In the face of a scarcity of experimental data (the *lhc* machine is not yet fully operational), we are forced to fancy our premature suggestions. It seems that the grand molecule may be akin to the founding ingredient $ML_n$ of *hts* oxocuprates and *cmr* manganates, so that a survey of its basic properties may be worth making. For that purpose we refer the reader to an earlier paper and the underlying literature [21]. In any event, the days ahead will be quite thrilling indeed and so will our quest for unraveling one of the top enigmas facing science today. Shall we? People of good faith are too sceptic. What do experimentalists say?